\documentclass[manuscript,preprint,authoryear,11pt]{aastex}

\usepackage[margin=1in]{geometry}
\usepackage{graphicx}
\usepackage{bm, amsmath, amssymb}
\usepackage[dvips]{color}

\newtheorem {remark} {Remark}

\newtheorem {theorem} {Theorem}

\begin{document}

\title{Numerical integration of a relativistic two-body problem via a  multiple scales method}

\author{Elbaz. I. Abouelmagd\altaffilmark{1,2,3}, S. M Elshaboury\altaffilmark{4}, H. H. Selim\altaffilmark{1}}

\altaffiltext{1}{Celestial Mechanics Unit - Astronomy Department - National Research Institute of Astronomy and Geophysics (NRIAG). Helwan - Cairo -  Egypt \newline
Email: eabouelmagd@gmail.com or eabouelmagd@nriag.sci.eg}

\altaffiltext{2}{Nonlinear Analysis and Applied Mathematics Research Group (NAAM) - Department of Mathematics - King Abdulaziz University.  Jeddah - Saudi Arabia }

\altaffiltext{3}{ Department of Mathematics - Faculty of Science and Arts - University of Jeddah.   Saudi Arabia }

\altaffiltext{4}{ Mathematics Department, Faculty of Science, Ain Shams University, Cairo, Egypt}

\begin{abstract}
 We offer  an analytical study on the dynamics of a  two-body problem perturbed by small post-Newtonian relativistic term.  We prove that, while the angular momentum is not conserved, the motion  is planar.  We also show that the energy is subject to small changes due to the relativistic effect. We also offer a periodic solution to this problem, obtained by a method based of separation of timescales. We demonstrate that our solution is more general than the method developed in the book by Brumberg (1991). The practical applicability of this model may be studies of the long-term evolution of relativistic binaries (neutron stars or black holes).
\end{abstract}


\section{Introduction}\label{s1}

The classical two-body problem is controlled by the interaction of two point masses moving under a mutual gravitational attraction and Newton's second law, where, the massive body is called the primary or the central body while the smaller body is called the secondary body. But in the framework of general relatively the mechanical laws and the equations of motion, which describe any dynamical system according to Einstein's theory are much more strenuous and complicated in analysis than under the assumptions of Newtonian mechanics. However the motion of celestial bodies under conceptions of Einstein mechanics  differ so little from their Newtonian representation. But for astronomical purposes, relativistic effects may be conveniently treated as a first-order perturbation. A comprehensive and extended study on relativistic celestial mechanics of the solar system, a theoretical development of gravitational physics as it applies to the dynamics of celestial bodies and the analysis of precise astronomical observations are presented by \cite{ko}.\\

  In a perturbed Keplerian orbits, the motion of a secondary body subjects to other forces in addition to the gravitational attraction of the central body. In this case, the other forces are called the perturbed forces. Which lead to a disturbance in the motion of the secondary body, however its effect may be very small comparison with the main forces. There are different kinds of perturbation forces, some of these decrease the velocity of the body as in drag forces, for more details \cite{sh} and  other increase the velocity as in thrust forces. There are also some forces cause a loss of orbital energy and angular momentum and coerce particles to slowly spiral to the sun in the case of our solar system as in the Poynting-Robertson (PR) effect and ion drag from the solar wind, see \cite{bu} for a comprehensive review of the various radiation forces.\\

 The perturbed forces can include:  Repulsive forces such as radiation pressure which depends on the cross section of the particle and other depends on the size, the electrical charge on a dust grain, which represented by Lorentz force.  A stochastic distributed dust between  a planet and the sun can be also considered as a perturbed force, this force generate supplementary random force on the orbiting particle, for instance \cite{je}, \cite{ma1},  \cite{ma2}, \cite{sha}. The non-sphericity effect , if one of the two bodies has irregular shape, the two-body problem becomes insoluble because the lack of sphericity in the body shape produce also an additional force, which can be also treated as a perturbed force, see \cite{j}, \cite{mar} and \cite{ab3}. Furthermore there are some interest papers which presented significant study on the effect of oblateness  in the framework of restricted three-body problem constructed by \cite{ab1} , \cite{abo} , \cite{ab3} , \cite{ab4} and  \cite{ab2} \\

Some of the aforementioned perturbed forces or one of them may be the only considerable perturbed force, if we assume that there are no other bodies outside the dynamical system. But this is not true for all real systems in space. For instant, Earth-Moon motion will suffer from an extra force of the gravitational attraction by the other planets and their satellites, this force is called the interaction force.
In most systems that involve multiple gravitational attractions, the dominating effect is produced by the central body. The central is a star in the case of stellar system and the other bodies are its planets, or a planet in the case of planetary systems and the other bodies are its satellites. The gravitational effect of the other body or bodies can be treated as a perturbation for unperturbed motion of the planet or satellite around its central body. But our main significant contribution  in the present paper is to  find the solution of the perturbed two-body problem in the framework of relativistic effect without secular terms by multiple scales method.\\

This paper is organized as follow: in Section \ref{s2} we briefly review  the $N-$body problem in the inertial and heliocentric references frames, as well as the perturbed acceleration of interaction gravitational forces.  In Section \ref{s3} we deduce the equations of motion of the perturbed two-body problem  by post-Newtonian  relativistic terms. In Sections \ref{s4}  we  analise  the  angular momentum and the energy relations.  In Section \ref{s5} we obtain the solution of the relativistic perturbed two-body problem by a multiple scales method. We end the paper with some  comments.

\section{Background}\label{s2}

\subsection{$N-$body problem in the inertial reference frame}

The classical $N-$body problem is defined in the framework of $N-$points moving under their gravitational attraction without other celestial bodies outside the system. Furthermore, we assume that $N=n+1$ such that each point mass $m_i$ is fully described by its mass $m_i$ and its position vector $\underline{x}_i(t)$ , $(i=0,1,2,.....,n+1)$ for all time $t$. In the case of our planetary system, $m_0$  may denote the mass of the Sun and it could be considered as the unit of mass. We may also define the velocity vector of each point mass by $\underline{\dot{x}}_i(t)$. If the initial state vectors  $\underline{x}_i(t_0)$ and $\underline{\dot{x}}_i(t_0)$ are known at the initial time $(t=t_0)$, then the task of finding the trajectories $\underline{x}_i(t)$ for each point mass $m_i$ in the inertial reference frame can be accomplished for all time $t$.\\

According to Newton's universal law of gravitation and his second law, we can write down the equations of motion in the inertial frame for The $N-$point masses in the form
\begin{equation}\label{eq01}
    \underline{\ddot{x}}_i=-G\sum_{j=0,j\neq i}^{n+1} m_j \dfrac{\underline{x}_{ji}}{x_{ji}^3}
\end{equation}
where $G$ is the constant of gravitation, $x_{ij}=|x_{ij}|=|x_{ji}|, \underline{x}_{ij}=x_{j}-x_{i}$ and $(i=0,1,2,.....,n+1)$.\\

The right hand side of Eq.(\ref{eq01}) represents the superpositions of gravitational forces acting on point mass $m_i$  under the condition that there are no masses outside the system of the $N-$point masses. It is important to note that Eq.(\ref{eq01}) represents an ordinary coupled nonlinear differential equations system of second order in time. The mass $m_i$ does not change with time which is not perfectly true in the most cases.

 \subsection{ $N-$body problem in the heliocentric reference frame}

The construction of Eq.(\ref{eq01}) is setup under the assumption that there are no masses out side of the system of the $N-$point masses which will never be $100\%$ true. We have to think in the gravitational attraction which the solar system experiences from our galaxy. If time periods of hundreds of millions of years are considered such effect must be taken into accounts (the revolution periodic of the solar system around the galactic center is estimated to be about $250$ million years, see for more details \cite{gb}. But if the mass $m_0$ dominates all other masses, this makes sense to rewrite the Eq.(\ref{eq01}) to describe the motion of the system relative to the point mass $m_0$. For that purpose, we define

\begin{equation}\label{eq02}
   \underline{r}_i(t)=\underline{x}_i(t)-\underline{x}_0(t)
\end{equation}
where $(i=1,2,.....,n+1)$ and the vector $\underline{r}_i$ is called the heliocentric position vector in our planetary system.\\

Starting from  Eq.(\ref{eq01}) with the help of Eq.(\ref{eq02}), we may easily set up the equation of motion for $N-$point masses in term of the heliocentric position vector $\underline{r}_i$ in the down form

\begin{equation}\label{eq03}
    \underline{\ddot{r}}_i=-G(m_0+m_i)\dfrac{\underline{r}_{i}}{r_{i}^3}-G\sum_{j=1,j\neq i}^{n+1} m_j \left [\dfrac{\underline{r}_{ji}}{r_{ji}^3}+\dfrac{\underline{r}_{j}}{r_{j}^3} \right]
\end{equation}
Eq.(\ref{eq03}) represents the equation of motion for $N-$point masses in the heliocentric coordinates system such that its origin follows the trajectory of the point mass $m_0$. This equation shows that the acceleration may be vanish in the heliocentric coordinates but that it is not possible for all time in inertial space according to Eq.(\ref{eq01}). In addition, the initial state vectors can be defined as $r_i(t_0)=r_{i0}$ and $\dot{r}_i(t_0)=\dot{r}_{i0}$ where $(i=1,2,.....,n+1)$.

\begin{remark}
 It is important to state that we are able to analyze the development of planetary system dynamics, without having defined the origin in the inertial space system relative to the central mass $m_0$, using Eq.\eqref{eq03} when the initial state vectors in the heliocentric system are given.
\end{remark}

\subsection{Equation of motion with the effect of interaction gravitational forces}

Starting from  Eq.\eqref{eq03} with the set up $m_{n+1}\equiv m$, $\underline{r}_{n+1}\equiv \underline{r}$ and $p$ denotes $n+1$,  without loss of the generality this equation can be written in the form
\begin{equation}\label{eq04}
    \underline{\ddot{r}}=-\mu\dfrac{\underline{r}}{r^3}+\underline{a}_p
\end{equation}
where $\mu=G(m_0+m)$ and $\underline{a}_p$ is the perturbed acceleration of the interaction force, which will be controlled by

\begin{equation}\label{eq05}
\underline{a}_p=-G\sum_{j=1}^{n} m_j \left [\dfrac{\underline{r}_{jp}}{r_{jp}^3}+\dfrac{\underline{r}_{j}}{r_{j}^3} \right]
\end{equation}

In Eq.\eqref{eq04} the first term on the right hand side represents the main term of the force which act on the point mass $m$ while the sum may be called the perturbation term, this property is correct especially in our planetary system, because the ratio of planetary or satellite masses to the mass of central body $(m_i/m_0\ll1)$ is small quantity, where the most massive planet, Jupiter,  has mass of $0.1 \%$ of the solar mass. Furthermore, one of the significant features of  Eq.\eqref{eq04}, the first term is a result of the Kepler force while  the second term is emerged from the interaction force. Actually, the last force is realistic and exist, because there are no encounters between the bodies for any real system. Thereby,  the perturbation due to interaction force is uniquely due to the transformation from the inertial reference frame to the heliocentric coordinates system.

\begin{remark}
 In the case of the point mass $m$ of negligible mass with respect to all other mass of the system, then $\mu=Gm_0$ and Eq.\eqref{eq04} enables us to describe the trajectory of a minor planet or a comet  in the heliocentric system.
\end{remark}

\section{Equations of motion with a relativistic effect}\label{s3}

The fixed body of spherical structure produce a spherical symmetric gravitational field with a metric in terms of  rectangular coordinates

\begin{equation}\label{eq198}
   ds^2=\left[\begin{array}{l} \left [p(r)c_l^2 dt^2 +2b(r)\dfrac{x^i}{r}c_l dtdx^i  \right] \\ \\
   -  \dfrac{1}{r^2} \left[a^2(r) \delta_{ik}+\left(q(r)-\dfrac{a^2(r)}{r^2}\right)x^i x^k \right]dx^idx^k
   \end{array}\right]
\end{equation}

where $c_l$ is the speed of light, in this setting $a, b, p$ and $q$ are are functions of $r$ to be determined from the field equations, $p$ and $b$ are arbitrary while $p$ and $q$ depend on $p$ and $b$. see \cite{ad}, \cite{br} and \cite{ko} for compleat details

In parameterized post-Newtonian approximation (PPN), the function $a(r)$ for most practically employed quasi-Galilean reference system may be controlled by
\begin{equation}\label{eq199}
  a(r)=r\left(1+(1-\alpha)\dfrac{m_c}{r}+\epsilon \dfrac{m_c^2}{r^2}+....\right)
\end{equation}
where $(\alpha , \epsilon , ...)$ are the parameters which characterize the type of coordinate, $m_c=\mathcal{G}M/c^2_l$ and $\mathcal{G}$ is the parameter which may differ from the universal gravitational constant by a constant factor such that $\mathcal{G}=\mathcal{A}G$, $M$ is the total mass of the system. For instance the values $\alpha=1, \epsilon=0$ regard to the standard coordinates, while harmonic coordinates are associated with $\alpha=0, \epsilon=0$, see\cite{br} for more classifications.

After inserting Eq.(\ref{eq199}) into Eq.(\ref{eq198}) the generalized Schwarzschild metric in the PPN approximation  take the following form in terms of rectangular coordinates
\begin{equation}\label{eq200}
   ds^2=\left[\begin{array}{l} \left[1-\dfrac{2m_c}{r}  +2(\beta-\alpha)\dfrac{ m_c^2}{r}+....\right]c_l^2 dt^2 \\ -\left[\delta_{ij}-\dfrac{2 m_c}{r}\left[(\gamma-\alpha)\delta_{ij} +\alpha \dfrac{x^ix^j}{r^2}\right] +....\right]dx^idx^j
   \end{array}\right]
\end{equation}
the parameters $\beta $ and $\gamma$ determine the features of the PPN formalism, for the general relativity $\beta=\gamma=1$. Nevertheless , considering $\beta$ and $\gamma$, the value of  $\alpha$ define specific coordinates conditions, with $\alpha=0$ the formula in Eq.(\ref{eq200}) reduces to the will-known Eddington Rebertson metric.

The rectangular coordinates $ x^1, x^2$ and $x^3$ related to the spherical coordinates $(r,\theta , \varphi )$ by
\begin{subequations}\begin{align}\label{eq201}
x^1&=r\cos \theta \sin \varphi\\
x^2&=r\sin \theta \sin \varphi\\
x^3&=r\cos \varphi \,
\end{align}\end{subequations}

Taking into account  the motion of two-body problem with a perturbation given by a relativistic effect, then Equation of relative motion in the type of vector form as described in \cite{br} and \cite{ko} will be governed by
\begin{subequations}\label{eq300} \begin{align}
&\underline{\ddot{r}}= -\frac {\mathcal{G}M}{r^3}\underline{r} + \underline{F}\\
&\underline{F}=\frac{\mathcal{G}M}{c^2_lr^3} \left[\left(2 \sigma\frac{\mathcal{G }M}{r} -2\epsilon \, \underline{\dot{r}}.\underline{\dot{r}}  + 3 \alpha \frac{(\underline{r}.\underline{\dot{r}})^2}{r^2} \right)\underline{r} +2\mu (\underline{r}.\underline{\dot{r}})\underline{\dot{r}} \right]
\end{align}\end{subequations}
where $\underline{r}=\underline{r}_2-\underline{r}_1$ is a relative vector, $\underline{r}_1$ ,  $\underline{r}_2$ are the position vectors of masses $M_1$ and $M_2$,  $M=M_1+M_2$. While $\alpha , \epsilon , \mu $ and $\sigma$ are arbitrary numerical parameters, and $\underline{F}$ is called the Chazy distributing force .
The above equation is an important  tool to investigate the motion of a test particle in the Schwarzschild gravitational field. The general relativity equations of the generalized Schwarzschild problem is a result for the setting:
\begin{subequations}\label{eq301}\begin{align}
\sigma&=\gamma+\beta-\alpha\\
2\epsilon&=\gamma+\alpha\\
\mu&=\gamma+1-\alpha\
\end{align}\end{subequations}

Substituting Eqs.\eqref{eq301} into Eqs.\eqref{eq300}, one obtain

\begin{equation}\label{eq302}
    \underline{\ddot{r}}= -\frac {\mathcal{G}M}{r^3}\underline{r}+\frac{\mathcal{G}M}{c^2_lr^3} \left[\left(2 (\gamma+\beta-\alpha)\frac{\mathcal{G }M}{r} -(\gamma+\alpha)\, \underline{\dot{r}}.\underline{\dot{r}}  + 3 \alpha \frac{(\underline{r}.\underline{\dot{r}})^2}{r^2} \right)\underline{r} +2(\gamma+1-\alpha) (\underline{r}.\underline{\dot{r}})\underline{\dot{r}} \right]
\end{equation}
or
\begin{subequations}\label{eq303}\begin{align}
&\ddot r - r \dot \theta^2 = -\frac{\mathcal{G}M}{r^2} +\frac{\mathcal{G}M}{c^2_lr^2} \left[2(\gamma+\beta-\alpha)\frac{\mathcal{G}M}{r}  - (\gamma+\alpha)r^2 \dot \theta^2 +  (\gamma+2)\dot r^2 \right] \\
&2 \dot r \dot \theta + r \ddot \theta =  \frac{2\mathcal{G}M(\gamma+1-\alpha)}{c^2_lr} \,\dot r\, \dot \theta
\end{align}\end{subequations}

Now we can simplify the equations by fixing  $\mathcal{G}M=\mathcal{G}(M_1+M_2)$ is equal unity and $\varepsilon=1/c^2_l$. For the general relativity, one of the choice is $\gamma=\beta=1$, furthermore $\alpha=0$ in harmonic coordinates. In this setting the parameter $\varepsilon$ represents the relativistic effect. Which its acceleration due to the theory of general relativity in maximum is of order $10^{-9}$ of the main term in the close Earth satellite. However this effect is very small , it is mandatory to take such effect into account for precise orbits determination. Hence  Eq.\eqref{eq302} and Eqs.\eqref{eq302} will be reduced to the form

\begin{equation}\label{eq304}
    \underline{\ddot{r}}= -\frac {1}{r^3}\underline{r}+\frac{\varepsilon}{r^3} \left[\left(\frac{4}{r} -\, \underline{\dot{r}}.\underline{\dot{r}}  \right)\underline{r} + 4(\underline{r}.\underline{\dot{r}})\underline{\dot{r}} \right]
\end{equation}
or
\begin{subequations}\label{eq305}\begin{align}
&\ddot r - r \dot \theta^2 = -\frac{1}{r^2} +\frac{\varepsilon}{r^2} \left[\frac{4}{r}  - r^2 \dot \theta^2 +  3\dot r^2 \right] \\
&2 \dot r \dot \theta + r \ddot \theta =  \frac{4\varepsilon}{r} \,\dot r\, \dot \theta
\end{align}\end{subequations}
The above equations are necessary and sufficient to investigate the structure of motion for a test particle in the Schwarzschild gravitational force in the framework of the parameterized post-Newtonian formalism.

\section{The constants of perturbed motion}\label{s4}
In this section we investigate how  the perturbation modifies the  angular momentum and the total energy.

\subsection{A modified conservation of the angular momentum and total energy}
\begin{theorem}\label{th01}
The angular momentum of the dynamical system for the perturbed two-body problem via the relativistic effect is not conserved, however its direction is a constant, while the total energy is conserved,  therefore the motion is a planar.
\end{theorem}
\textbf{\emph{Proof Theorem 1}}

Let $\underline{h}_{t}$ be the vector of total orbital angular momentum per unit mass of the body $m$ according to the heliocentric reference frame, therefore
\begin{equation}\label{eq09}
\underline{h}_t = \underline{r} \wedge \underline{\dot{r}}=|\underline{r} \wedge \underline{\dot{r}}|\underline{\hat{h}}_t
\end{equation}
where  $h_t=|\underline{r} \wedge \underline{\dot{r}}|$ and $\underline{\hat{h}}_t$ is the unit normal vector of the orbital plane.\\

  Multiply the equation of motion, Eq.\eqref{eq304} with the vectorial product by $\underline{r}$, one obtain
\begin{equation}\label{eq10}
   \underline{r} \wedge \underline{\ddot{r}}=\dfrac{4\varepsilon \dot{r}}{ r^2} \underline{r} \wedge \underline{\dot{r}}
\end{equation}
now we can rewrite Eq.\eqref{eq10} in the below form
 \begin{equation}\label{eq11}
 \dot{\underline{h}}_t =\dfrac{4\varepsilon \dot{r}} {r^2} \underline{h}_t
\end{equation}
taking the scalar product of $\underline{h}_t$ with Eq.\eqref{eq11}
\begin{equation}\label{eq11a}
 \dot{\underline{h}}_t \wedge  \underline{h}_t=0
\end{equation}
Eq.\eqref{eq11a} admits that the direction of the angular momentum is a constant therefore the motion is a planer.

In the framework of the first parameterize post-Newtonian approximation the Lagrangian will be written in the following form, see \cite{da} and \cite{bl}

 \begin{equation}\label{eq600}
    L = L_N + \dfrac{1}{c^2_l}L_{1PN}
\end{equation}
therefore under  the aforementioned setting in Sec.(\ref{s3}), $L_N$ and $L_{1PN}$ are given by
\begin{subequations}\label{eq601}\begin{align}
L_N&=\dfrac{1}{2} \underline{\dot{r}}^2 + \dfrac{1}{r}\\
L_{1PN}&= \dfrac{1}{2r} \left[\dfrac{r}{4} +(\underline{\dot{r}}^2)^2 +3 \underline{\dot{r}}^2-\dfrac{1}{r} \right]
\end{align}\end{subequations}
substituting Eqs.(\ref{eq601}) int Eq.(\ref{eq600}) the $1PN$ Lagrangian in the harmonic coordinates is of the type

\begin{equation}\label{eq603}
 L=\dfrac{1}{2} v^2 +\dfrac{1}{r}+   \dfrac{\varepsilon}{2r}\left [ \dfrac{r}{4} (v^2+ 3\underline{\dot{r}}^2-\dfrac{1}{r} \right]
\end{equation}
where $v^2= \underline{\dot{r}}^2$

 Consequently Eq.(\ref{eq603}) admit the constants of motion as
  \begin{subequations}\begin{align}\label{eq604}
\underline{h}&=\underline{r}\wedge  \dfrac{\partial L}{\partial \underline{\dot{r}}} \\
E&= \underline{\dot{r}}.\dfrac{\partial L}{\partial \underline{\dot{r}}} -L
\end{align}\end{subequations}
or
\begin{equation}\label{eq605}
 \underline{h}=\underline{r} \wedge \underline{\dot{r}} \left [1+ \dfrac{1}{2}\varepsilon \left(v^2+ \dfrac{6}{r}\right)\right]
\end{equation}

\begin{equation}\label{eq606}
 E=\dfrac{1}{2} v^2 -\dfrac{1}{r}+  \dfrac{\varepsilon}{2r}\left [ \dfrac{3r}{4} v^4+ 3v^2+\dfrac{1}{r} \right]
\end{equation}

 Since the total angular momentum $h_t=|\underline{r} \wedge \underline{\dot{r}}|$, then we can rewrite Eq.(\ref{eq605}) in the form
\begin{equation}\label{eq607}
 h_t=h\left [1- \dfrac{1}{2}\varepsilon \left(v^2+ \dfrac{6}{r}\right)\right]
\end{equation}

from  Eq.(\ref{eq606}), we can rewrite  Eq.(\ref{eq607}) in the form
\begin{equation}\label{eq608}
 h_t=h\left [1- \varepsilon \left(\dfrac{4}{r}+E\right)\right]
\end{equation}
We would like to refer that  Eqs.(\ref{eq606} - \ref{eq608}) are considered direct results for the provided relations in the book by Brumberg (1991), for case of general relativity ($\gamma=\beta=1$ and $\alpha=0$). Specifically, our equations (\ref{eq605}) and (\ref{eq606}) correspond to the expressions (3.1.52) and (3.1.53) in \cite{br}, accordingly.
%\textcolor{red}{ ~$\longleftarrow$ PLEASE CHECK IF THIS IS CORRECT}

Eq.(\ref{eq608}) shows that  the magnitude of $h_t$  is a function in the radial vector $r$, this implies that the total angular momentum is not conserved. Since $\underline{h}_t=\underline{r}\wedge{\underline{\dot{r}}}$, then we conclude  that the angular momentum vector $\underline{h}_t$ is always perpendicular to the plane (a plane of motion) which includes both vectors $\underline{r}$ and $\underline{\dot{r}}$. This prove that the position and velocity vectors lie in the plane of motion and the total angular momentum $h_t$ is not conserved as in the classical case.
\begin{remark}
 The change in the magnitude of the angular momentum is very small due to the perturbation of the general relativity, because the term of the relativistic perturbation depends on the factor $1/c^2$. So if this term is neglected the total angular momentum is conserved  and the structures of the orbital plane of motion coincide with the unperturbed model.
\end{remark}

Finally, we state that $E$ represent the total energy which is conserved, while the angular momentum is not.

In this context, the mechanical energy-like invariant of motion  is given by

 \begin{equation}\label{eq40}
   \dfrac{1}{2}\bar{v}^2-\dfrac{1}{r}+\dfrac{7\varepsilon}{ r^2}=E_l
 \end{equation}
 where $\bar{v}^2=\left(1-\dfrac{6 \varepsilon}{r}\right)v^2$ and $E_l$ is a constant representing the total energy, see \emph{\textbf{Appendix}} for details.
 \begin{theorem}\label{th02}
 In neglect of $O(\varepsilon e)$ in Eq.(\ref{eq40}), the energy-like approximate integral of motion $E_l$ becomes an exact integral of motion as in expression of Eq.(\ref{eqenergy}).
 \end{theorem}
 \begin{equation}\label{eqenergy}
   \dfrac{1}{2}{v}^2-\dfrac{1}{r}=E
\end{equation}
here $E=E_l-4\varepsilon/h^4$

 The proof of this statement is available in the \emph{\textbf{Appendix}}

\emph{Theorem 2} establishes that the energy-like approximate integral of motion coincides with the appropriate energy emerging in the unperturbed motion. The difference between the approximate integral of motion and the exact one is of the order of $O(\varepsilon)$

\begin{theorem}\label{th03}
For low-eccentricity orbits, the total angular momentum of the relativistically perturbed two-body problem  is conserved.
\end{theorem}
\textbf{\emph{Proof Theorem 3}}\\

Since
\begin{equation}\label{eq500}
    \underline{r}.\underline{\dot{r}}=r\dot{r}
\end{equation}
then the scalar product $\underline{r}.\underline{\dot{r}}$ can be approximated in the form
\begin{equation}\label{eq501}
    \underline{r}.\underline{\dot{r}}=-h\left[e-e^2 \cos f +e^3 \cos^2f +O(e^4)\right]
\end{equation}
for low eccentricity, the scalar product $\underline{r}.\underline{\dot{r}}$ is in the order  of $O(e)$. Then we may state that $\underline{r}.\underline{\dot{r}}\approx 0$ . Furthermore  the energy theorem of two-body problem may be reduced for the low of eccentricity orbit to $\underline{\dot{r}}.\underline{\dot{r}}\approx 1/r$. In this context, we can write Eq.(\ref{eq304}) in the form

\begin{equation}\label{eq15}
    \underline{\ddot{r}}=-\left[\dfrac{1}{r^3}-\dfrac{3 \varepsilon }{ r^4} \right]\underline{r}
\end{equation}
Now we take the vector product of $\underline{r}$ with Eq.(\ref{eq15}), one obtain
\begin{equation}\label{eq16}
   \ddot{\underline{r}}\wedge \underline{r}=0
\end{equation}
  After integration, the above equation can be written as
\begin{equation}\label{eq17}
  r^2 \dot{\theta}=h
\end{equation}
where $h$ is a constant.  This shows that  the total angular momentum is conserved as in the classical case without any changes in its magnitude.

\section{Solution provided by multiple scales method}\label{s5}
\hskip 0.5cm

We start by rewriting Eqs.(\ref{eq305}) in the  form

\begin{equation}\label{eq41}
    \ddot{r} - \dfrac{h^2_t}{r^3} = - \dfrac{1}{r^2} + \varepsilon \left(\dfrac{4}{r^3} - \dfrac{h^2_t}{r^4} +\dfrac{3 \dot{r}^2}{r^2}\right)
\end{equation}

Letting $r = \dfrac{1}{u}$,  we obtain

\begin{equation}\label{eq42}
   \dfrac{d^2u}{d\theta^2} + u = \dfrac{1}{h^2} +  \varepsilon\left(a_0+a_1u+u^2+(\dfrac{du}{d\theta})^2\right)
\end{equation}
where $a_0=2E/h^2$ and $a_1=4/h^2$

Eq.(\ref{eq42}) is considered a more general than the well-known Binet's equation, see for instant \cite{nr}

\subsection{Solution of unperturbed problem}

Eq.(\ref{eq42}) represents the trajectory of the second body around the primary. It is important to note that this trajectory will follow a Kepler's orbit when the effect of general relatively is switched off ($\varepsilon = 0$). Therefore, the solution is given by
\begin{equation}\label{eq43}
    u(f)= \dfrac{1}{h^2} \, (1 + e \, \cos f)
\end{equation}
where $f$ is the true anomaly, $e= \kappa\,h^2$ is the orbit eccentricity and $\kappa$ is a constant to be determined from the initial conditions
\begin{equation}\label{eq44}
    u (0) = u_0 = \dfrac{1}{h^2}(1 + e)\qquad \text{and} \qquad  \dfrac{d}{df}u(0)= 0
\end{equation}
furthermore Eq.(\ref{eq43}) represents a periodic solution where $u(f + 2\pi) = u(f)$.\\

Now we want to find a solution of Eq.(\ref{eq42}) with the effect of perturbation parameter $\varepsilon$. So we have to use some perturbation methods. For instance, KBM method, Lindstedt-Poincar\'e technique and the method of multiple scales or the classical theory of perturbation. The first two method provide a way to obtain asymptotic approximations of periodic solutions, see for more details \cite{ce} and \cite{ab3}. But they cannot be used to find solutions that form aperiodically on a slow variable scale. The method of multiple scales method is a more general approach in which we construct a solution includes one or more new slow variables for each interest parameter scale in the problem. It does not require that this solution depends periodically on the slow variable.

\subsection{Multiple scales method}

 A solution obtained by the classical perturbation approach will include a secular term leading to  unbounded growth.  The multiple scales  method eliminates such unwanted secularities . The analysis procedure of multiple scales method will be established in the following four steps:
 \bigskip

\begin{itemize}

\item  \textbf{ The first step}\\

We look for a solution $u(f)\equiv y_n(f,\tau)$ where $\tau:=\varepsilon f$. Here $y_n$ depends on  two variables scales, namely $f$ is  a fast variable  and $\tau$ is a slow variable  such that $\tau$ is not negligible when $f$ is of order $O(\varepsilon^{-1})$. We can extend this procedure to many variables scales as we like, but in this case the other variables scales will have order  $O(\varepsilon^n)$ , $(n=2,3,...)$. Consequently we will obtain more equations which add more difficulty without any further insight into the method, especially with a very small value for $\varepsilon$. So here we consider two variables scales only.\\

\item \textbf{ The second step}\\

 We write $u$ as   a perturbation series:

 \begin{equation}\label{eq45}
    u (f) = \sum\limits_{n = 0}^\infty \epsilon^n y_n(f,\tau)
\end{equation}
Thereby, the solution of Eq. (\ref{eq42}) can be given by $u(f)$, which depends only on the  variable $f$. Nevertheless the multiple scales method seeks solution that is a function of both the fast true variable scale $f$ and the slow variable $\tau$. Even though in the actual solution $f$ and $\tau$ are in correlation to each other, this method treats them as independent variables. This strategy enables us to eliminate the secular effect by elegant way. We want to emphasize that $f$ and $\tau$ are ultimately not independent. Now we  expand the notation of the derivative with respect to $f$ by a differential operator $D_f$:

\begin{equation}\label{eq46}
   \dfrac{d}{df} u (f) = D_f\sum\limits_{n = 0}^\infty \epsilon^n y_n(f,\tau)
\end{equation}
for all $f$, $\dfrac{d}{df}\tau=\varepsilon$ and the formula of the differential operator $D_f$ is given by
\begin{equation}\label{eq47}
   D_f:=(\dfrac{\partial}{\partial f}+ \varepsilon \dfrac{\partial}{\partial \tau})
\end{equation}

We now inserting Eq.(\ref{eq47}) into Eq.(\ref{eq46}) and assume that $y_n$ is continuously differentiable with respect to $f$ and $\tau$. The first and second derivatives of $u$ are given by
\begin{subequations}\begin{align}
    \dfrac{du}{df}=\dfrac{\partial y_0}{\partial f} + \varepsilon \left(\dfrac{\partial y_0}{\partial \tau}+ \dfrac{\partial y_1}{\partial f} \right) + \varepsilon^2 \left(\dfrac{\partial y_1}{\partial \tau}+ \dfrac{\partial y_2}{\partial f} \right) +O(\varepsilon^3) \qquad  \qquad \qquad  \qquad \label{eq48a}\\
     \dfrac{d^2u}{df^2}=\dfrac{\partial^2 y_0}{\partial f^2} + \varepsilon \left(\dfrac{\partial^2 y_1}{\partial f^2}+ 2\dfrac{\partial^2 y_0}{\partial \tau \partial f} \right) + \varepsilon^2 \left( \dfrac{\partial^2 y_2}{\partial f^2}+ 2\dfrac{\partial^2 y_1}{\partial \tau \partial f}   + \dfrac{\partial^2 y_0}{\partial \tau^2} \right) +O(\varepsilon^3) \, \, \label{eq48b}
\end{align}\end{subequations}
where$\dfrac{d}{d\theta}=\dfrac{d}{df}$.\\

\item \textbf{ The third step}\\

 In third step we  find $y_n$ with the condition in Eq.(\ref{eq44}). Therefore we insert Eqs.(\ref{eq48a},(\ref{eq48b}) and Eq.(\ref{eq45}) into Eq.(\ref{eq42}) and comparing the terms in orders of $\varepsilon$, we obtain a sequence of linear partial differential equations where the first three of this sequence are

\begin{equation}\label{eq49}
\begin{array}{l}
 \dfrac{\partial^2y_0}{\partial f^2}+y_0=\dfrac{1}{h^2} \, \\
     y_0(0,0) =\dfrac{1}{h^2}\left(1+e \right) , \qquad  \dfrac{\partial}{\partial f}y_0(0,0)=0
\end{array}
\end{equation}

\begin{equation}\label{eq50}
\begin{array}{l}
  \dfrac{\partial^2y_1}{\partial f^2} + y_1 =a_0+a_1 y_0+y_0^2+\left(\dfrac{\partial y_0}{\partial f}\right)^2 -\dfrac{\partial^2y_0}{\partial\tau \partial f}\, \, \\
     y_1(0,0) =0 \,, \qquad  \dfrac{\partial}{\partial f}y_1(0,0)=-\dfrac{\partial}{\partial \tau}y_0(0,0)
\end{array}
\end{equation}

\begin{equation}\label{eq51}
\begin{array}{l}
  \dfrac{\partial^2y_2}{\partial f^2} + y_2 =a_1y_1+2y_0 y_1 +2 \dfrac{\partial y_0}{\partial f}\left(\dfrac{\partial y_0}{\partial \tau}+\dfrac{\partial y_1}{\partial f}\right)-2\dfrac{\partial^2y_1}{\partial \tau \partial f} -\dfrac{\partial^2y_0}{\partial \tau^2} \, \, \\
     y_2(0,0) =0 \,, \qquad  \dfrac{\partial}{\partial f}y_2(0,0)=-\dfrac{\partial}{\partial \tau}y_1(0,0)\,.
\end{array}
\end{equation}

 \item \textbf{ The fourth step}\\

 In this step we calculate the solutions of the previous equations by eliminating the secular terms. It is convenient to assume that the solution of Eq.(\ref{eq49}) can be written in the following form to obtain the dependence of the solutions due to the pervious partial differential equations in the independently treated variables $f$ and $\tau$

 \begin{equation}\label{eq52}
   y_0 (f,\tau) = \dfrac{1}{h^2}+A(\tau)e^{if}+\bar{A}(\tau)e^{-if}
\end{equation}
where $A(\tau)$ is shall be a yet arbitrary complex function of $\tau$ and $\bar{A}(\tau)$ denotes its complex conjugate. Furthermore these functions will be determined with the conditions that the solutions of $y_1(f,\tau)$ has no secular terms.

Now substituting Eq.(\ref{eq52}) and its partial derivative with respect to $f$ into  Eq.(\ref{eq50}), one obtain

\begin{equation}\label{eq53}
\begin{array}{l}
  \dfrac{\partial^2y_1}{\partial f^2} + y_1 =b_0 +4 A(\tau)\bar{A}(\tau) +s_{1}(\tau)e^{if}+\bar{s}_{1}e^{-if} \, \, \\
     y_1(0,0) =0 \,, \qquad  \dfrac{\partial}{\partial f}y_1(0,0)=-\dfrac{\partial}{\partial \tau}y_0(0,0)
\end{array}
\end{equation}
where $b_0=\dfrac{1}{h^2}(2E+\dfrac{5}{h^2})$ and
\begin{equation}\label{eq54}
    s_{1}(\tau)=b_1 A -2\dfrac{dA}{d\tau}i
\end{equation}
\begin{equation}\label{eq55}
    \bar{s}_{1}(\tau)=b_1 \bar{A}+\dfrac{d\bar{A}}{d\tau}i
\end{equation}
where $b_1=6/h^2$

Notice that  $e^{\pm if}$ is a solution of the homogeneous equation associated to Eq.(\ref{eq50}). Therefore if the coefficients $s_{1}(\tau)$ and $\bar{s}_{1}(\tau)$ are nonzero, then the solution of $y_1(f,\tau)$ will include secular terms in the variable $f$. But that is exactly what we want to avoid. Hence we set  $s_{1}$ and $\bar{s}_{1}$ equal to zero. In addition Eq.(\ref{eq55}) is just the complex conjugate of Eq.(\ref{eq54}), then it can be omitted. If $A(\tau)$ satisfies the conditions  $s_{1}(\tau)=0$ and $\bar{s}_{1}(\tau)=0$, the solution of $y_1$ will not contains secular terms and at least no secularities appear in the first two terms in the perturbation series.

To achieve our objective, let us try to solve Eq.(\ref{eq54}) in the framework of the polar coordinates $(R,f)$, so we assume that
\begin{equation}\label{eq56}
\begin{array}{l}
 A(\tau)=R(\tau)e^{if}  \\
 R \, , f:\mathbb{R}\rightarrow \mathbb{R}
\end{array}
\end{equation}
substituting Eq.(\ref{eq56}) into Eq.(\ref{eq54}) with the condition  $s_{1}(\tau)=0$, with simple calculations one get
\begin{subequations}\label{eq57}\begin{align}
 A(\tau)=R_0e^{i(f_0-c\tau)}\label{eq57a}\\
  \bar{A}(\tau)=R_0e^{-i(f_0-c\tau)}\label{eq57b}
\end{align}\end{subequations}
where $c=3/h^2$ , $R_0$ and $f_0$ are arbitrary constants.\\

Substituting Eq.(\ref{eq57}) into Eq.(\ref{eq52}) and using the initial conditions in  Eq.(\ref{eq49}), one obtain $R_0= e/2h^2$ and $f_0=0$. Hence Eq.(\ref{eq52}) can be rewritten in the form
\begin{equation}\label{eq58}
   y_0 (f,\tau) = \dfrac{1}{h^2}+R_0\left (e^{i(f-c\tau)}+e^{-i(f-c\tau)}\right)
\end{equation}

\end{itemize}

 The steps are completed with construction of Eq.(\ref{eq58}), but we have to note that   Eqs.(\ref{eq57}) and Eq.(\ref{eq58}) investigate that  Eq.(\ref{eq53}) is in resonance with the solution of its homogenous part and its general solution will contain secular terms. To avoid this secularities we must find the general solution of $y_1(f,\tau)$ with the conditions that the partial differential equation due to $y_2(f,\tau)$ has no secular terms. Consequently we can do the same procedure to determine the other functions of the perturbed series.

 Given $y_0$ as in (\ref{eq58}), the  solution $u (f,\tau) $ can be written in the below form

\begin{equation}\label{eq59}
   u (f,\tau) = \dfrac{1}{h^2} \left[1+e \cos (f-c\tau)\right]
\end{equation}
 or
\begin{equation}\label{eq60}
   u (f) = \dfrac{1}{h^2}\left[1+ e\cos (1-\varepsilon c)f\right]
\end{equation}

where $e$ is the eccentricity when $\varepsilon=0$

In this setting the error for fixed $f$ is given by
\begin{equation}\label{eq61}
   err (u) := |u(f)-y_0(f)|
\end{equation}
is at most of order $\textit{ O }(\varepsilon)$. But this is only true as long as $f< \varepsilon^{-1}$. Therefore $y_1$ stays bounded for all $f, \tau$, hence $\varepsilon y_1\in \textit{O}(\varepsilon)$ and we only have to worry about $\varepsilon^2 y_2 $ which we did not investigate. This term may contain a secular terms that grows $\textit{O}(f)$. That is why we have to set the validity interval to $f \in [0,f_0/\varepsilon]$ for some $f_0 >0$, if the term of $y_1$ is included. According to Eq.(\ref{eq61}) the relative error may be $\in \left[0\,,2e/(1+e)\right)$

We have to compar the solution obtained by multiple scales method directly with analytical or semi-analytical known solution
, of course with a predefined accuracy.  For this purpose we will rewrite Eq.(\ref{eq60}) in the following form

\begin{equation}\label{eq700}
   r = \dfrac{a\left(1-e^2\right)}{\left[1+ e\cos \left(1-\dfrac{3 \, \varepsilon}{a \left(1-e^2\right)}\right)f\right]}\quad.
\end{equation}
In the framework of the parameterize post-Newtonian approximation when  ($\gamma=\beta=1$ and $\alpha=0$ ) in the case of general relativity,  Eq.(\ref{eq700}) is deduced from the provided relations in the book by Brumberg's book,  correspond to the expressions (3.1.60 - 3.1.65) in Brumberg (1991).

 However the expression of Eq.(\ref{eq700}) is considered a zero-th order approximation of multiple scales method,  We have revealed the timescale separation underlying the Brumberg old result. On the other hand, Brumberg's solution was based on a template borrowed from the unperturbed two-body problem~---~an anzats that we do not use. So we can conclude that the multiple scales method is more general than the Brumberg method. In fact, the latter method provides an approximation to the former.

 \section{Conclusions}
 In this work, we investigate the two-body problem perturbed by a post-Newtonian relativistic term. In Theorem 1, we show that the motion is planar and the total energy is conserved while the angular momentum is not. In Theorem 2, we prove that the like-invariant of motion can be reduced to invariant motion. In Theorem 3, we offer the conservation law of the angular momentum for the low eccentricity orbits. We also offered a method based on multiple time scales, to obtain an approximated solution to the relativistic perturbed two-body problem. We have demonstrated that our approach is more general than the one developed in the book by Brumberg (1991). The latter method furnishes
 an approximation to our solution. Thereby, we have revealed the timescale separation underlying Brumberg's old result.

 \section*{Appendix}\label{s7}

 \subsection*{Mechanical energy-like invariant of motion}

  Here we derive an energy-like integral of motion, by taking the scalar product of Eq.(\ref{eq304}) with $\lambda \dot{\underline{r}}$ instead of $\dot{\underline{r}}$,where $\lambda =\lambda (t)$ is a scalar function of time which can be establish of the time of the invariant integral. Thus we obtain
\begin{center}
\begin{equation*}
\lambda \ddot{\underline{r}}.\dot{\underline{r}}=-\left[\dfrac{1}{r^2}-\dfrac{ \varepsilon }{ r^3}\left(4+3v^2r \right) \right]\lambda \dot{r}  \qquad \qquad \qquad \qquad \qquad \qquad \qquad
A.1
\end{equation*}
\end{center}
the above equation can be rewritten in the form
\begin{equation*}
 \dfrac{1}{2}\dfrac{d}{dt}(\lambda v^2)=-\left(\dfrac{1}{r^2}-\dfrac{ 4 \varepsilon}{ r^3}\right)\lambda\dot{r} + \left(\dfrac{1}{2}\dot{\lambda} +\dfrac{ 3 \varepsilon \lambda \dot{r}}{ r^2} \right)v^2\qquad \qquad \qquad \qquad
 A.2
\end{equation*}
In order to obtain the integration of Eq.\,(A.2) in closed form, we must eliminate the second term in the right hand side of this equation. This implies that
\begin{center}
\begin{equation*}
 \dfrac{1}{2}\dot{\lambda} +\dfrac{ 3 \varepsilon \lambda \dot{r}}{ r^2}=0 \qquad \qquad \qquad \qquad \qquad \qquad \qquad \qquad \qquad \qquad \qquad
 A.3
\end{equation*}
\end{center}
after integration of Eq.\,(A.3), the function $\lambda$ is given by
\begin{equation*}\label{eq37}
\lambda(t)=A e^{-6\varepsilon/r} \qquad \qquad \qquad \qquad \qquad \qquad \qquad \qquad \qquad \qquad \qquad
A.4
\end{equation*}
where $A$ is an arbitrary constant of integration dose not equal zero. Since $\varepsilon$ is a very small quantity, then we can expand the right hand side of Eq.\,(A.4) and restrict ourself with the first order of $\varepsilon$. therefore this equation can be rewritten in the form
\begin{equation*}
\lambda(t)=A (1-\dfrac{6\varepsilon}{r}) \qquad \qquad \qquad \qquad \qquad \qquad \qquad \qquad \qquad \qquad
A.5
\end{equation*}
substituting Eq.\,(A.5)  into Eq.\,(A.2) and integration with neglect the terms of $O(\varepsilon^2)$ or more, we obtain
\begin{equation*}
   \dfrac{1}{2}\lambda v^2-\dfrac{A}{r}+\dfrac{7\varepsilon A}{ r^2}=\bar{E}\qquad \qquad \qquad \qquad \qquad \qquad \qquad \qquad \qquad
   A.6
\end{equation*}
 $\bar{E}$ is the integration constant  and $\bar{v}^2=(1-\dfrac{6 \varepsilon}{r})v^2$. Thus Eq.\,(A.6) will be take the new form
\begin{equation*}
   \dfrac{1}{2}\bar{v}^2-\dfrac{1}{r}+\dfrac{7\varepsilon}{ r^2}=E_l \qquad \qquad \qquad \qquad\qquad \qquad \qquad \qquad \qquad \qquad
   A.7
\end{equation*}
It is clear that Eq.\,(A.7) represent an expression of the energy-like invariant integral with an extra term which characterize the relativistic effect where $E_l=\bar{E}/A$.

\subsection*{Proof Theorem 2}

Since  $\bar{v}^2/r$ and $1/r^2$ can be written in below form
\begin{subequations}\begin{align*}
\dfrac{\bar{v}^2}{r}&=\dfrac{1}{2}v^2-\dfrac{3\varepsilon}{h^4}\left[1+3e \cos f+e^2 \left(1+s \cos^2 f\right)+e^3\sin^2 f \cos f\right]  \qquad A.8\\
\dfrac{1}{r^2}&=\dfrac{1}{h^4}\left[1+2\cos f +e^2 \cos f\right] \qquad \qquad \qquad \qquad\qquad \qquad \qquad \qquad A.9\,
\end{align*}\end{subequations}
substituting Eqs.\,(A.8 , A.9) into Eq.\,(A.7), after neglecting the terms with coefficient $O(\varepsilon e)$. One obtain
\begin{equation*}
   \dfrac{1}{2}{v}^2-\dfrac{1}{r}+\dfrac{4 \varepsilon}{ h^4}\left(1+\dfrac{5}{4}e \cos f -\dfrac{1}{ 4}e^2 (3-\cos^2 f)-\dfrac{3}{4}e^3 \cos f\right)=E_l \qquad
   A.10
\end{equation*}
with  neglecting terms of coefficient $O(\varepsilon e)$,  Eq.\,(A.10) is reduced to
\begin{equation*}
   \dfrac{1}{2}{v}^2-\dfrac{1}{r}=E \qquad \qquad \qquad \qquad\qquad \qquad \qquad \qquad \qquad \qquad \qquad \qquad
   A.11
\end{equation*}
where $E=E_l-4\varepsilon/h^4$

Hence the energy-like approximate integral of motion becomes as in  the unperturbed motion.

To investigate the accuracy of the like-invariant motion, it must be compered with the exact invariant motion. Let $err(E)=E-E_l$ be the  difference between both constants of motion, one obtain
\begin{equation*}
  err(E)=\varepsilon \left[\dfrac{3}{8}v^4 +\dfrac{9}{2r}v^2-\dfrac{13}{2r^2}\right]\qquad \qquad \qquad \qquad\qquad \qquad \qquad \qquad  \qquad \qquad
  A.12
\end{equation*}
According to  Eq.\,(A.12), we can conclude that this error will vanish when $v^2=\pm v^2_r$, where $v^2_r= 2(2\sqrt{30}-9)/3r$, decreases for $v^2 \in (-v^2_r\,, v^2_r)$ and increases for $v^2 \in (-\infty\,, -v^2_r)\cup (v^2_r\,, \infty)$. On the other hand, Eq.\,(A.12) can be written in the form
\begin{equation*}
  err(E)=-\dfrac{13 \varepsilon}{8h^4} \left[1-\dfrac{16}{13}e \cos f- \dfrac{42}{13} e^2 \left(1+\dfrac{16}{21} \cos^2 f\right)-\dfrac{48}{13} e^3 \cos f -\dfrac{3}{13}e^4\right]\qquad
  A.13
\end{equation*}
Consequently , the deviation of the energy-like approximate integral of motion from the exact integral is of order of $O(\varepsilon)$

 \subsection*{Acknowledgments}

 The authors would like to thank Cristina Stoica for several constructive suggestions. The authors are also grateful to the referees, in particular to Michael Efroimsky, for the careful reading of the manuscript and their creative suggestions, which have lead to the improvement of the quality and the clarity of the present work.

\end{document}